\documentclass[a4paper,fleqn,usenatbib,useAMS,letter]{mnras}
\usepackage[usenames,dvipsnames]{color}
\usepackage[normalem]{ulem}

\usepackage{graphicx}	
\usepackage{amsmath}	
\usepackage{amssymb}	
\usepackage{multicol}        
\usepackage{bm}		
\usepackage{pdflscape}	

\usepackage[T1]{fontenc}
\usepackage{ae,aecompl}

\usepackage{newtxtext,newtxmath}

\title[Convective-reactive nucleosynthesis of K, Sc, Cl and p-process isotopes in O-C shell mergers]{Convective-reactive nucleosynthesis of K, Sc, Cl and p-process isotopes in O-C shell mergers}

\author[C. Ritter et al.]{
C. Ritter$^{1,6,7}$\thanks{Contact e-mail: \href{mailto:critter@uvic.ca}{critter@uvic.ca}},
R. Andrassy$^{1,6}$,
B. C{\^o}t{\'e}$^{1,2,6,7}$, 
F. Herwig$^{1,6,7}$,
P. R. Woodward$^{3,6}$,
\newauthor 
M. Pignatari$^{4,7}$,
S. Jones$^{5,7}$ \\
\\
$^{1}$Department of Physics and Astronomy, University of Victoria, Victoria, BC, V8P5C2, Canada \\
$^{2}$National Superconducting Cyclotron Laboratory, Michigan State University, MI, 48823, USA \\
$^{3}$LCSE and Department of Astronomy, University of Minnesota, Minneapolis, MN 55455, USA \\
$^{4}$E. A. Milne Centre for Astrophysics, Department of Physics \& Mathematics, University of Hull, HU6 7RX, United Kingdom \\
$^{5}$Heidelberg Institute for Theoretical Studies, Schloss-Wolfsbrunnenweg 35, D-69118 Heidelberg, Germany \\
$^{6}$JINA-CEE, Michigan State University, East Lansing, MI, 48823, USA \\
$^{7}$NuGrid collaboration, http://www.nugridstars.org
}

\date{Last updated 2016 May 22; in original form 2013 September 5}

\pubyear{2016}

\newlength{\apjcolwidth}
\newlength{\figwidth}
\newlength{\doublewide}
\newcommand{\fig}[1]{Fig.\,\ref{#1}}
\newcommand{\ppmstar}{\textsc{PPMstar}}
\newcommand{\mppnp}{\textsc{mppnp}}

\newcommand{\omegac}{\textsc{OMEGA}}
\newcommand{\msun}{\ensuremath{\, {\rm M}_\odot}}
\newcommand{\lsun}{\ensuremath{\, {\rm L}_\odot}}

\usepackage{isotope}

\begin{document}

\label{firstpage}
\pagerange{\pageref{firstpage}--\pageref{lastpage}}
\maketitle

\begin{abstract}
We address the deficiency of odd-Z elements P, Cl, K and Sc in
galactic chemical evolution models through an investigation of the
nucleosynthesis of interacting convective O- and C shells in massive stars.
3D hydrodynamic simulations of O-shell
convection with moderate C-ingestion rates show no dramatic deviation
from spherical symmetry. We derive a spherically
averaged diffusion coefficient for 1D nucleosynthesis
simulations which show that such convective-reactive ingestion events can be a
production site for P, Cl, K and Sc.
An entrainment rate of $10^{-3}\msun$/s features overproduction factors $OP_\mathrm{s} \approx 7$.
Full O-C shell mergers in our 1D stellar
evolution massive star models have overproduction factors $OP_\mathrm{m}>1 \mathrm{dex}$
but for such cases 3D hydrodynamic simulations suggest deviations from spherical symmetry.
$\gamma$-process species can be produced with overproduction factors of $OP_\mathrm{m}>1 \mathrm{dex}$, e.g.\ for
\isotope[130,132]{Ba}.
Using the uncertain prediction of the $15\msun$, $Z=0.02$ massive
star model ($OP_\mathrm{m} \approx 15$) as representative for merger
or entrainment convective-reactive events involving O- and C-burning
shells, and assume that such events occur in more than 50\% of all stars, our chemical evolution models reproduce the observed Galactic trends of the odd-Z elements.
\end{abstract}

\begin{keywords}
stars: abundances - stars: evolution - stars: interiors - hydrodynamics - galaxy: abundances
\end{keywords}

\section{Introduction}\label{s.introduction}

Massive stars are the main producers of intermediate-mass elements below the Fe peak \citep[e.g.][]{woosley:02}. 
However, the odd-Z elements K and Sc in galactic chemical evolution (GCE) models of the Milky Way based on yields of \citet[][K06]{kobayashi:06} or \cite{nomoto:13} are up to 1 dex lower compared to halo and disk stars. Several promising
production scenarios have been considered for Sc, such as the ejecta of proton-rich neutrino winds \citep{froehlich:06}, jet-induced core-collapse supernova explosions \citep[e.g.][]{tominaga:09}
and hypernovae \citep[e.g.][]{sneden:16}. However the impact of these mechanisms on GCE models has not yet been demonstrated.

Massive stars are also considered the main source of rare p-rich stable isotopes
beyond iron, although the underproduction in simulations
compared to the solar abundances has always been recognized \citep[see][and references therein]{pignatari:16c}.

While previous attempts to explain the production of odd-Z
elements focussed on explosive nucleosynthesis environments, we are
proposing that odd-Z elements, as well as possibly p-process species, are forming in convective-reactive events
in pre-supernova stellar evolution. 
In 1D stellar models the mixing-length theory \citep{cox:68} describes convection in a time and spatially averaged way  that provides realistic results in conditions where the nuclear reaction time scale is much larger than the convective mixing timescale, i.e.\ the Damk\"ohler number
$Da = \tau_\mathrm{mix}/\tau_\mathrm{reac} << 1  $.
In that case the convective region is instantaneously mixed compared to the reaction time scale and the detailed shape of the mixing profile is not important. However, in convective-reactive nucleosynthesis $Da \sim 1$ and the combined mixing and nuclear processes at each radial location determines the overall nuclear production of the region. In more extreme cases the energy feedback from nuclear reactions on the convective turn-over time scale may be dynamically relevant. This situation can be encountered in the late phases of stellar evolution of both low-mass and massive stars, such as  in simulations of combustion of H-rich material ingested into He-shell flash
convection in post-AGB stars. This can lead to a 3D non-radial instability Global Oscillation of Shell H-ingestion \citep[GOSH,][]{herwig:14}.
This convective-reactive environment gives rise to exotic i-process nucleosynthesis \citep{herwig:11,denissenkov:17}.

2D hydrodynamic simulations of \citet{meakin:06} suggested that the 
entrainment at the top of the O shell in a simultaneous  O- and C-shell convection simulation can reach $10^{-4}\msun/s$ and significantly affect the evolution.  Nucleosynthesis in 1D models of O-C shell mergers
has been mentioned in the literature
\citep{rauscher:02,tur:07}. In the NuGrid model library  \citep[][in prep., R17]{ritter:17} O-C shell mergers are found 
in stellar models with initial mass between $12\msun$ and $20\msun$ at $Z=0.02$
and $Z=0.01$ (Table \ref{tab:runs}).

\begin{table}
\begin{center}
\resizebox{\columnwidth}{!}{%
\begin{tabular}{lrrrrrr}
\hline 
Run ID & \multicolumn{6}{l}{Hydrodynamic simulations} \\
\hline
&  $\dot{M}_e$ [$\msun$/s] &$L_{O}$  [$\lsun$]& $L_{C}$ [$\lsun$] & $f_{QO}$ & $f_{QC}$& Stationary \\
\hline
I2  &  $3.13\times10^{-7}$ & $4.27\times 10^{10}$ &  $8.29\times 10^{9}$ & 1.0 & 1.0& yes \\
I13 & $1.07\times10^{-4}$ & $5.10\times 10^{12}$ &  $2.54\times 10^{12}$ & 67.5 & 1.0 & yes \\
I11 & $2.24\times10^{-4}$ & $8.02\times 10^{11}$ & $1.32\times 10^{13}$ & 13.5 & 10.0 & no \\
\hline
Run ID & \multicolumn{6}{l}{1D post-processing simulations of I2} \\
\hline
&  $\dot{M}_e$  [$\msun$/s] & $\overline{OP_\mathrm{s}}$  & $f_D$ & $t_{tot}$ [min] & & \\
\hline
Sm5  & $1.2\times 10^{-5}$ & 1.23 & 1.0 & 16.5 & &  \\
Sm5L  & $1.2\times 10^{-5}$ & 1.65  & 1.0 & 291.7 & &  \\
Sm4  & $1.2\times 10^{-4}$ & 2.38  & 1.0 & 16.5 & &  \\
Sm3  & $1.2\times 10^{-3}$ & 4.70   & 1.0 & 16.5 & & \\
Sm3D  & $1.2\times 10^{-3}$ &  7.00 &15.7  & 16.5  & & \\
\hline
Run ID & \multicolumn{6}{l}{Stellar evolution tracks} \\
\hline
& $\overline{OP_\mathrm{m}}$  & $M_{\mathrm{ini}}$ [$\msun$] & Z & Reference & & \\
\hline
M15Z0.02 & 14.74 & 15 & 0.02 & R17 &  & \\
M12Z0.01 & 8.34  & 12 & 0.01 & R17  &  & \\
M15Z0.01 & 33.53 & 15 & 0.01 & R17 &   &\\
M20Z0.01 & 12.15 & 20 & 0.01 & R17 &   &\\
M25Z0.02R &  & 20 & 0.02 & R17 & &\\
M25Z0.02J &  & 25 & 0.02 & J17 &  & \\
\end{tabular}
}
\caption[]{
Overview of simulations and their properties. Given are the entrainment rates $\dot{M}_e$, time-averaged O- and C-burning luminosities $L_{O}$ and $L_{C}$,
the increase of the Q values of O-shell and C-shell fluid burning by factors $f_{QO}$ and $f_{QC}$ and
an indication whether the entrainment process can be considered stationary. 
The luminosities of I13 and I11 vary in the time interval adopted for averaging by up to a factor of $\sim 4$. The entrainment rate in I11 was derived using a different method to be described in \citep{andrassy:17} and corresponds to the last 2.5\,min before the flow became so viollent that the simulation had to be stopped.
Stellar evolution tracks of R17 with mean overproduction factors of convective O-C shell mergers $\overline{OP}$ of P, Cl, P and Sc, the increase of the diffusion coefficient
by a factor $f_D$ and the total run time $t_{tot}$.
M$_{\mathrm{ini}}$ and Z are the initial mass and metallicity of the stellar evolution models.}
\label{tab:runs}
\end{center}
\end{table}

We explore C ingestion into a convective O-shell
with 3D hydrodynamic simulations. A diffusion coefficient profile
is derived and applied in 1D nucleosynthesis simulations.
Shell mergers in stellar evolution models are analyzed and predictions tested
against observations using GCE models.
The simulation tools are introduced in Section \ref{s.methods}, the results are in Section \ref{s.results} and
discussion and conclusion in Section \ref{s.discussion}.

\section{Methods}\label{s.methods}

We perform 3D simulations of the first convective O shell as in \citet[][J17]{jones:17}. We
adopt the same radial stratification based on a $25\msun$ stellar evolution
model (M25Z0.02J, Table \ref{tab:runs}) and the same numerical approach  \citep[3D hydrodynamics code
\ppmstar,][]{woodward:15}. We run simulations in 4$\pi$ geometry on a $768$$^3$ grid of the ingestion of C-rich material from a stable layer atop the O shell. Instead of driving the convection with a constant volume heating term, we now use a realistic O burning prescription according to Eq.\,18.75 in \cite{kippenhahn:12}. 

Our initial nucleosynthesis analysis of
C-shell material ingestion into the O shell with a 1-zone simulation
showed that the
\isotope[12]{C}(\isotope[12]{C},$\alpha$)\isotope[20]{Ne}
\citep{caughlan:88} and the
\isotope[16]{O}($\alpha$,g)\isotope[20]{Ne} reactions produce most of
the energy over most of the O shell. We assume that each $\alpha$ particle liberated
by the first reaction immediately triggers the second since plenty of
\isotope[16]{O} is available. Thus, the energy release of C burning is
the sum of Q values of the two reactions.  The mass fraction of
\isotope[12]{C} in the C shell of the stellar model M25Z0.02J is
0.026. We use five times that value for the fluid in the top stable
layer in the \textsc{PPMstar} simulations to shorten the simulations'
transition to a stationary state, which reduces the overall
computational costs. In addition to this simulation (I2) we perform
additional runs with enhanced O- and C-burning energy release
(Table~\ref{tab:runs}).

The \isotope[16]{O} + \isotope[12]{C} reaction, which we ignore
  in this initial set of hydro simulations, may also become
  relevant, sensitively depending on the
  temperature, the mixing efficiency and the amount of entrained
  C-shell material. Indeed, in the I2 (I13) hydro simulations the C-shell
  fluid is mixed fast to the hottest layers in quantities such that the
  reaction provides a factor 8 (0.7) more energy than the reactions included in the hydro network.
  A detailed analysis of this complex, non-linear problem is beyond the scope of this work and will be
  addressed elsewhere (Andrassy et al. in prep).

The detailed nucleosynthesis is computed with the 1D multi-zone
post-processing code \mppnp\ as in \cite{herwig:11}. The
stratification is the same as in the hydrodynamic simulation I2, which
is based on the stellar model M25Z0.02J.  C-shell material with an
abundance distribution of the post-processed stellar model M25Z0.02R
at a range of rates is injected into the upper part of the convective
shell.  Mixing is modelled based on a diffusion
coefficient determined from the 3D simulations I2
(Fig.~\ref{fig:mppnp_ingestion_setup}).

We model the chemical evolution of the Milky Way with a one-zone,
open-box model with galactic inflows and outflows in
\omegac\ \citep{cote:16b,cote:17} which is part of the NuGrid chemical evolution framework
at \url{http://nugrid.github.io/NuPyCEE}.
Here we use yields of AGB and
massive star models from NuGrid  \citep[][in prep.]{pignatari:16,ritter:17}, and alternatively for massive stars from K06, as well as 
for PopIII stars \citep{heger:10} and SNIa 
\citep{seitenzahl:13}.  The initial mass function of 
\citet{kroupa:01} is adopted over the initial mass range from $0.1\msun$ to
$100\msun$, and we assume the ejection of stellar yields between the
initial masses of $1\msun$ and $30\msun$.

\section{Results}\label{s.results} 


\subsection{Convection and feedback in 3D}	
\label{convfeedback}

3D hydrodynamic simulations of J17 of the first O shell show a mass
entrainment rate of $1.3\times10^{-6}\msun$/s.
If the O shell of that underlying stellar model grows outward at this rate 
it can reach the C shell in 1.9 days. This is before the end of O shell convection which suggests the
possibility of significantly enhanced entrainment rates or even an O-C shell merger.

\begin{figure}
\begin{center}
\includegraphics[width=0.9\columnwidth]{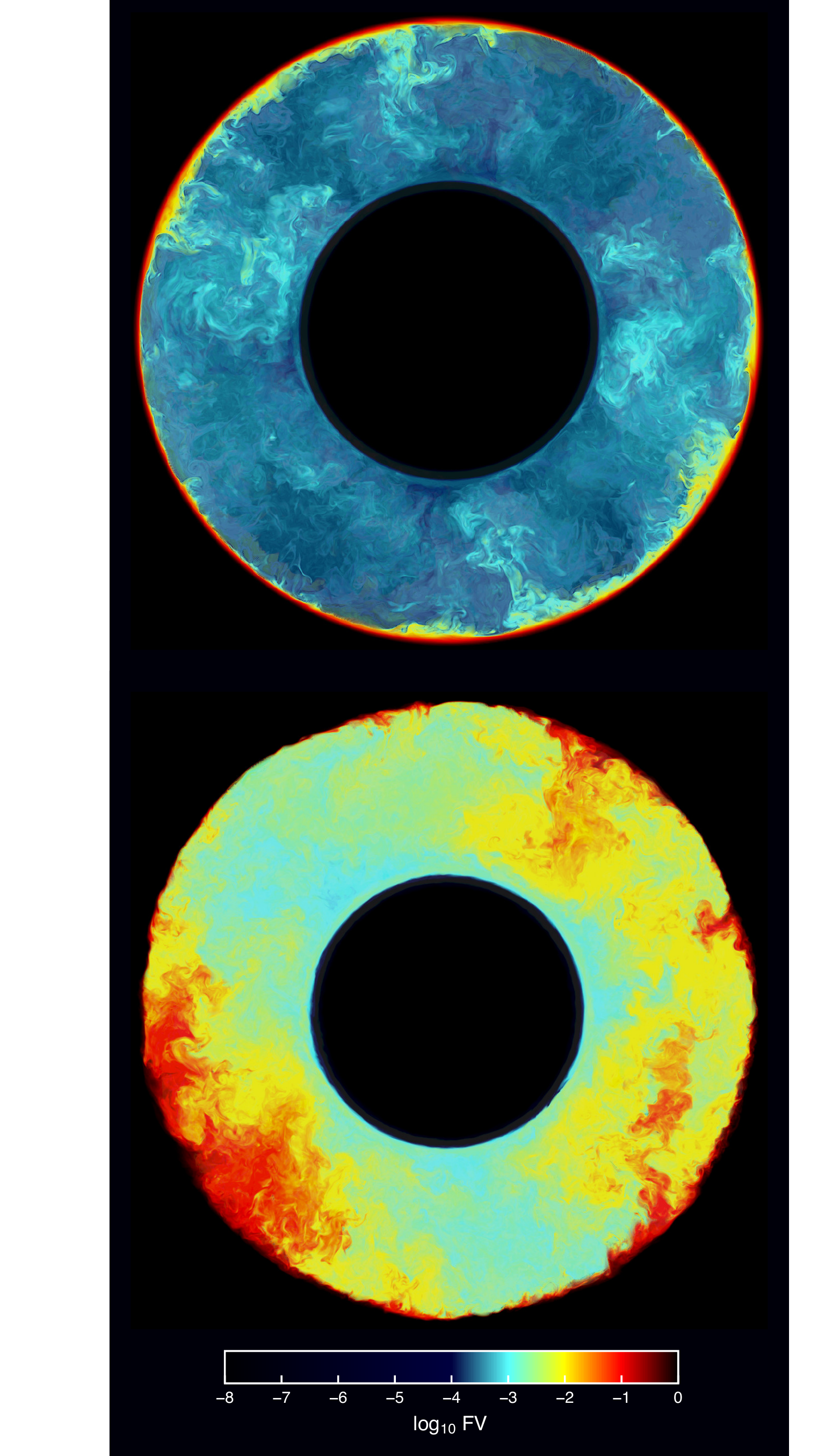}
\end{center}
\caption{Volume fraction of the C-rich fluid in a sphere slice of 
  3D convective O-shell simulations on a $768^3$ grid. \emph{Top:} Simulation I2 after $148\mathrm{min}$.
  \emph{Bottom:} Run I13 after $10.5 \mathrm{min}$. The entrainment rate is 342 times
  higher than in I2.}
\label{fig:hydro_rendering}
\end{figure}
In simulation I2, C-rich material is entrained into the O shell over 110 convective turn-over times of $132 \mathrm{s}$. The burning of the entrained material is turned off in the code for the first 34 turnovers.
The entrained fluid reaches the bottom of the convective O shell (\fig{fig:hydro_rendering}) and, after C burning has been turned on, its spherically averaged abundance develops towards 
a stationary state with an entrainment rate of $3.13\times10^{-7}\msun$/s. 
This rate is lower than that measured by J17 because (1) the luminosity is lower in our case and
(2) our experience to date with the \ppmstar code's present version indicates a decline in ingestion rate at very low
luminosities that falls below well established trends that we observe at higher luminosities. We suspect this fall-off
to be due to numerical rather than physical causes at the grid resolution of run I2, namely $768^3$ cells.
Checking that the ingestion rate has converged using simulations of double the grid resolution of run I2 is quite
expensive, and such checks have been successfully carried out already for high luminosity cases (cf. J17). 
We are addressing this issue instead by modifying the code to compute in future a potentially fully nonlinear 
perturbation to the star's very rapidly varying base state.

The energy release from C burning does not significantly affect the flow properties in I2.
The stationary nature and approximate spherical symmetry of the convective shell justify approximating the 3D mixing with a diffusion coefficient (see J17 for details) and applying it in 1D nucleosynthesis models.

Significantly higher ingestion rates might occur when the O-shell merges with the C-shell, or in 
a later and more luminous pre-supernova O shell \citep[J17,][]{meakin:06}. For this phase the 1D stellar models of R17
do show in fact O-C shell mergers for several cases (Table \ref{tab:runs}). In order to start exploring 3D hydrodynamic properties of such high-entrainment or even merger regimes we artificially increase the O-burning luminosity in simulation I13 by $\approx 2 \mathrm{dex}$ to $5.1\times10^{12}\lsun$ by increasing the Q value of the O-burning reaction by a factor of $67.5$, which corresponds to the increase between the first and second O shell of the stellar model M15Z0.02 (Table \ref{tab:runs}).
The average entrainment rate is $1.07\times10^{-4}\msun$/s in this case, in
agreement with the entrainment-luminosity law of J17. The flow is more
inhomogeneous and the convective boundary is significantly more deformed than in
I2 (\fig{fig:hydro_rendering}), but we do not find any global entrainment
instability \citep[c.f.][]{herwig:14}. Therefore, at this high entrainment rate nucleosynthesis may still be estimated from spherically symmetric \mppnp\ simulations. 

Although the mass exchange in shell mergers could initially be
hindered by the entropy gradient between the shells it has to be
considered that strong non-spherical instabilities similar to the GOSH
can occur with an unkown range of effective C-burning luminosities. In order to explore such scenarios we perform as an experiment simulation I11 (Table \ref{tab:runs}) were we enhance the
energy release from O and C burning fluid by factors of $13.5$ and
$10.0$ respectively. In this
case a violent, global, non-radial oscillation does emerge, in which case 
the 1D approach would break down. 

\subsection{Nucleosynthesis in 1D}	
\begin{figure}
\centering
\includegraphics[width=0.8\columnwidth]{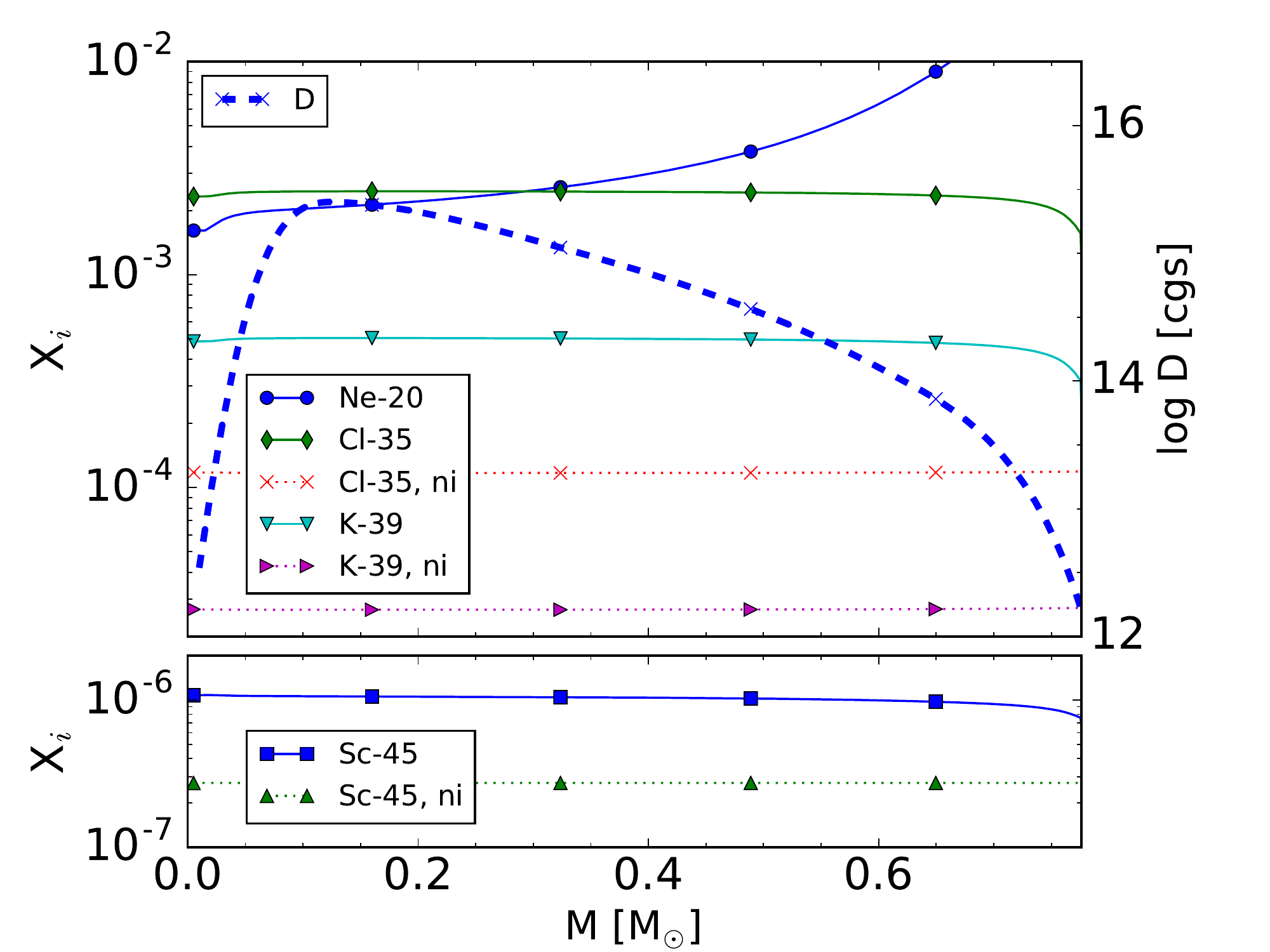}
\caption{Abundance profiles of run Sm4 after 16.5 min of ingesting C-shell material at a rate of  $1.2\times10^{-4}\msun$/s.
For comparison, the abundances produced over the same time with no ingestion (ni).
The diffusion coefficient D is extracted from the 3D hydrodynamic simulation I2.}
\label{fig:mppnp_ingestion_setup}
\end{figure}	

We perfom 1D \mppnp\ simulations for a range of ingestion rates (Table
\ref{tab:runs}), from $1.2\times10^{-7}\msun$/s to
$1.2\times10^{-3}\msun$/s, at which point the whole C
shell would be entrained in $1000 \mathrm{s}$ which is approximately
eight overturning timescales. In the limit of near-sonic mass
transport the mass exchange rate could even reach $1\msun$/s.

With entrainment rates corresponding to hydro run I2 no relevant
production of odd-Z elements is observed. For an entrainment rate of
$1.2\times10^{-5}\msun$/s the O shell may reach the C shell in 291.7min, without
any significant element production (Sm5,Sm5L). For entrainment
rates that correspond to 3D simulation I13 (\mppnp\ run Sm4, Table
\ref{tab:runs}) and above considerable amounts of \isotope[35]{Cl},
\isotope[39]{K} and \isotope[45]{Sc} (\fig{fig:mppnp_ingestion_setup})
as well as \isotope[31]{P} are produced in the lower part of the O
shell through convective-reactive entrainment and burning of
\isotope[20]{Ne}.
The element production of Cl, K and Sc is larger for higher convective velocities and for higher entrainment rates (Table
\ref{tab:runs}).

Through repeated multi-zone simulations with individual rates turned
off it emerges that burning of ingested Ne leads to the production of
odd-Z isotopes \isotope[31]{P}, \isotope[35]{Cl}, \isotope[39]{K} and
\isotope[45]{Sc} at different depths in the convective zone.
($\gamma,$p) reactions, such as \isotope[31]{P}($\gamma$,p), release
protons which produce odd-Z elements through reactions such as
\isotope[38]{Ar}(p,$\gamma$)\isotope[39]{K}. A further detailed
investigation of this convective-reactive nucleosynthesis site is
required to fully determine the nucleosynthesis paths that involve
conditions at several layers simultaneously.

\begin{figure}
\centering
\includegraphics[width=\columnwidth]{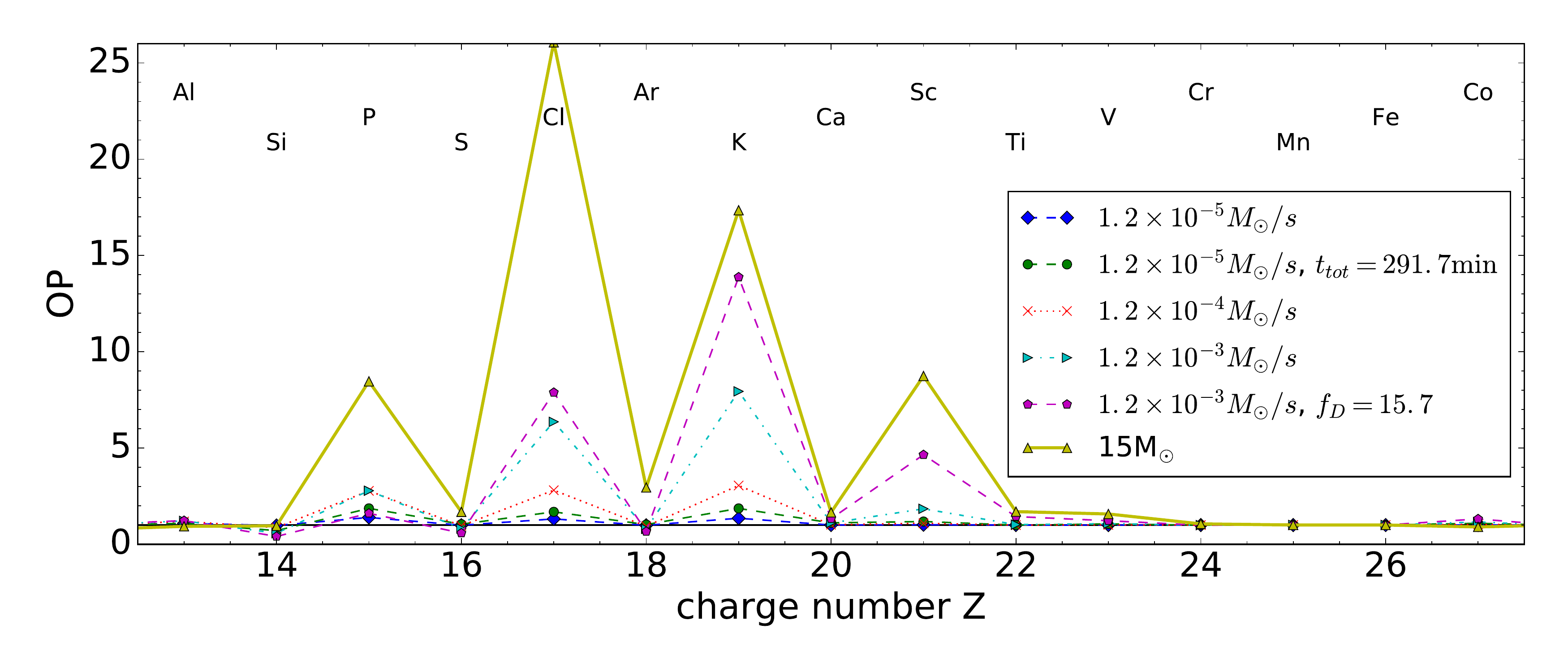}
\caption{
  Overproduction factors OP$_\mathrm{s}$ due to C-shell ingestion for
  different entrainment rates of C-shell material, total run times and
  diffusion coefficients in our synthetic models (Table
  \ref{tab:runs}). In comparison the overproduction factors
  OP$_\mathrm{m}$ in the O-C shell merger of the stellar model
  M15Z0.02.
  }
\label{fig:shellmerger_prodfacs_combined}
\end{figure}

In order to consider properly contributions from all burning shells
to the total stellar yields we compare the material
produced in an O-shell with ingestion ($Y_{i}$) to the material that
would result without ingestion ($Y_{ni}$) relative to the total yields
$Y_{tot}$ of the stellar model M25Z0.02R
(\fig{fig:shellmerger_prodfacs_combined}).  Thus, the overproduction
factor for 1D \mppnp\ runs is $OP_\mathrm{s} = (Y_{i} + Y_{tot})/
(Y_{ni} + Y_{tot})$.

\begin{figure}
\centering
\includegraphics[width=0.8\columnwidth]{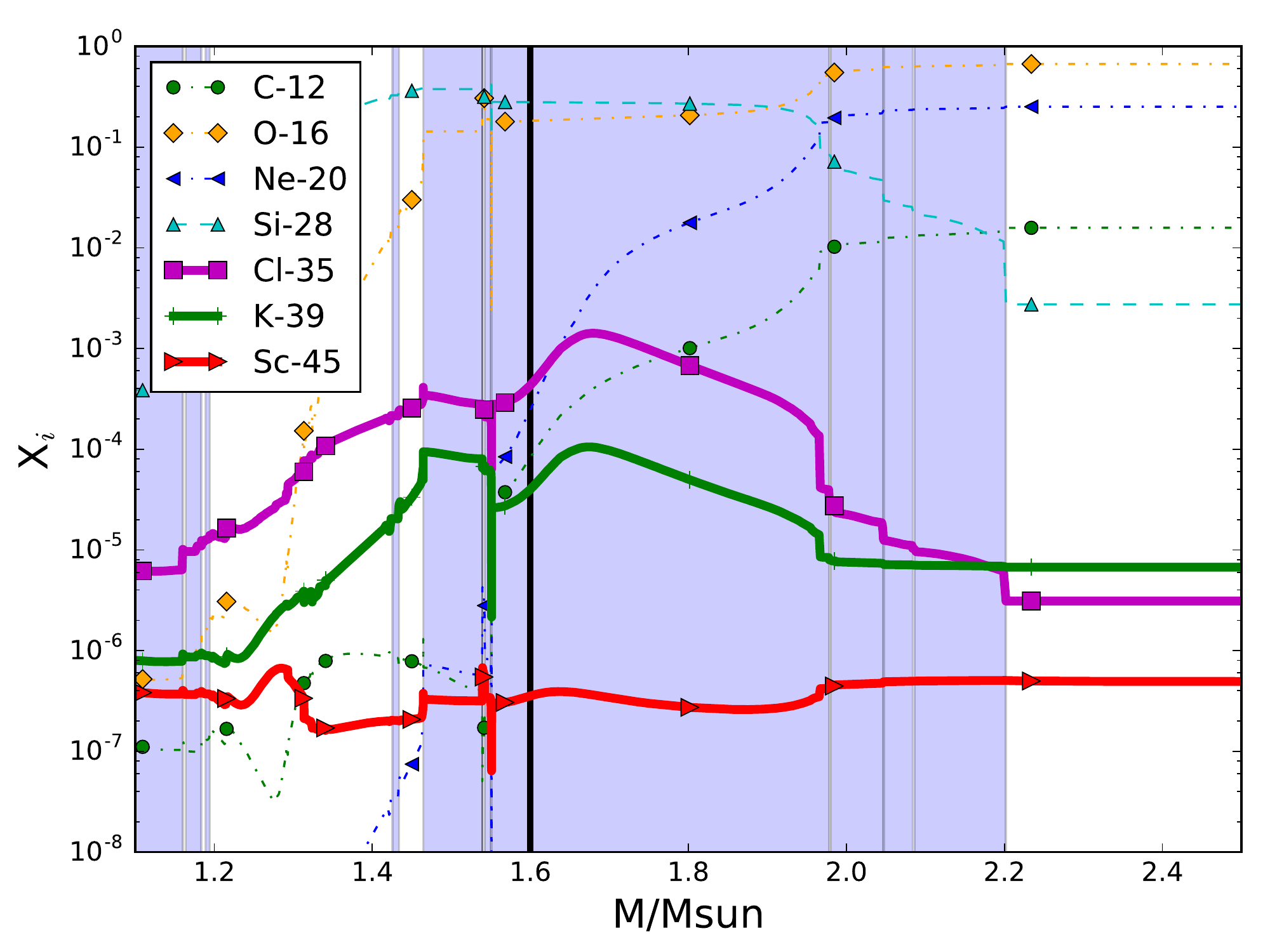}
\caption{Abundance profiles during entrainment of C-shell material into the O shell about 4 min after the end of convective Si core burning in model M15Z0.02. The blue shades indicate convective regions. The black vertical line shows the position of the remnant mass of the CCSN explosion model.} 
\label{fig:shellmerger_profile_mesa}
\end{figure}

The 1D stellar model M15Z0.02 experiences an O-C shell merger about $4
\mathrm{min}$ after the end of convective Si core burning. After an
initial partial mixing phase both convective shells merge
(\fig{fig:shellmerger_profile_mesa}), and large amounts of
\isotope[35]{Cl}, \isotope[39]{K} and \isotope[45]{Sc} are produced
(\fig{fig:shellmerger_prodfacs_combined}). Again, the 1D simulations
are not reliable at such high entrainment rates, but merely indicative
(see Section \ref{convfeedback}).
For stellar models the overproduction factors are calculated as $OP_\mathrm{m} = Y_{f}/Y_{ini}$ where $Y_{ini}$ and $Y_{f}$ are the total amounts of material above the mass cut at onset and after inter-shell mixing, respectively (Table
\ref{tab:runs}). 
Fallback is taken into account with the adoption of the mass cut of the delayed explosion prescription as in R17. 
The overproduction factors due to the O-C shell merger in the stellar
model are qualitatively similar to what we found in our 1D \mppnp\ ingestion simulations (\fig{fig:shellmerger_prodfacs_combined}).

Due to its deep location, the O shell is significantly modified by the core-collapse SN explosion and affected by fallback.
We use the CCSN prescription for the stellar model M15Z0.02 with a neutron-star remnant coordinate
based on the delayed explosion prescription of \cite{fryer:12}.  
The remnant coordinate is below the peak production of  \isotope[31]{P}, \isotope[35]{Cl}, \isotope[39]{K} and \isotope[45]{Sc} (\fig{fig:shellmerger_profile_mesa}).
The overproduction factors based solely on the explosive nucleosynthesis 
indicate that P, Cl, K and Sc are little affected in our explosion model. 


The $\gamma$ process occurs in Ne and O shell burning in the CCSN explosion
of massive star models through photo-disintegration reactions on heavy elements \citep{woosley:78,pignatari:16c}.
During an O-C shell merger, ``fresh'' heavy elements are constantly transported down to the O-burning shell, providing new seeds for photo-disintegration.

\begin{figure}
\centering
\includegraphics[width=\columnwidth]{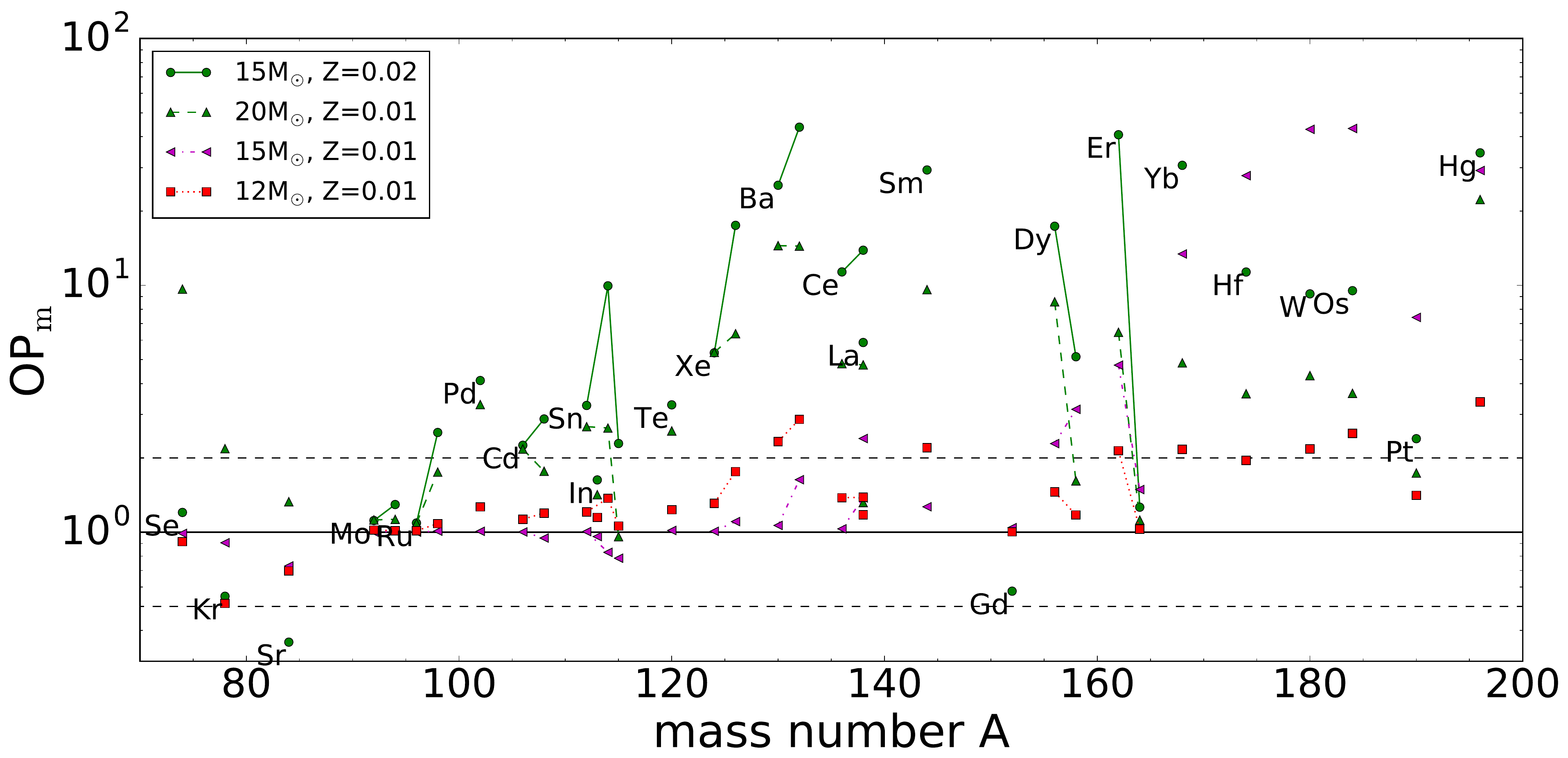}
\caption{Overproduction factors OP$_\mathrm{m}$ of p nuclei in O-C shell mergers of
  stellar models of NuGrid.  \isotope[180]{Ta} values are omitted for
  clarity and are 2.95, 0.7, 5.61 and 0.64 in legend order, starting
  from the top.  The dashed horizontal lines show an overproduction
  factor of 0.5 and 2.}
\label{fig:shellmerger_prodfacs_hydro_ppr}
\end{figure}

We compare the overproduction factors of the classical 35 p nuclei of
the O-C shell merger in the stellar models M15Z0.02, M20Z0.01,
M15Z0.01 and M12Z0.01 (\fig{fig:shellmerger_prodfacs_hydro_ppr}).
In these models we find strong variations in the overproduction
factors with stellar mass and metallicity.  Light p nuclei are
destroyed while heavier species are effectively produced. Most of
those species have the largest overproduction factors in the stellar
model M15Z0.02.  We confirm the results of \cite{rauscher:02} that p
nuclei can be made in an O-C shell merger.  At this level of production,
the impact of O-C shell mergers may change the GCE scenario of at least
some of the p nuclei.


\subsection{K and Sc trends in the Milky Way}

\begin{figure}
\centering
\includegraphics[width=0.8\columnwidth]{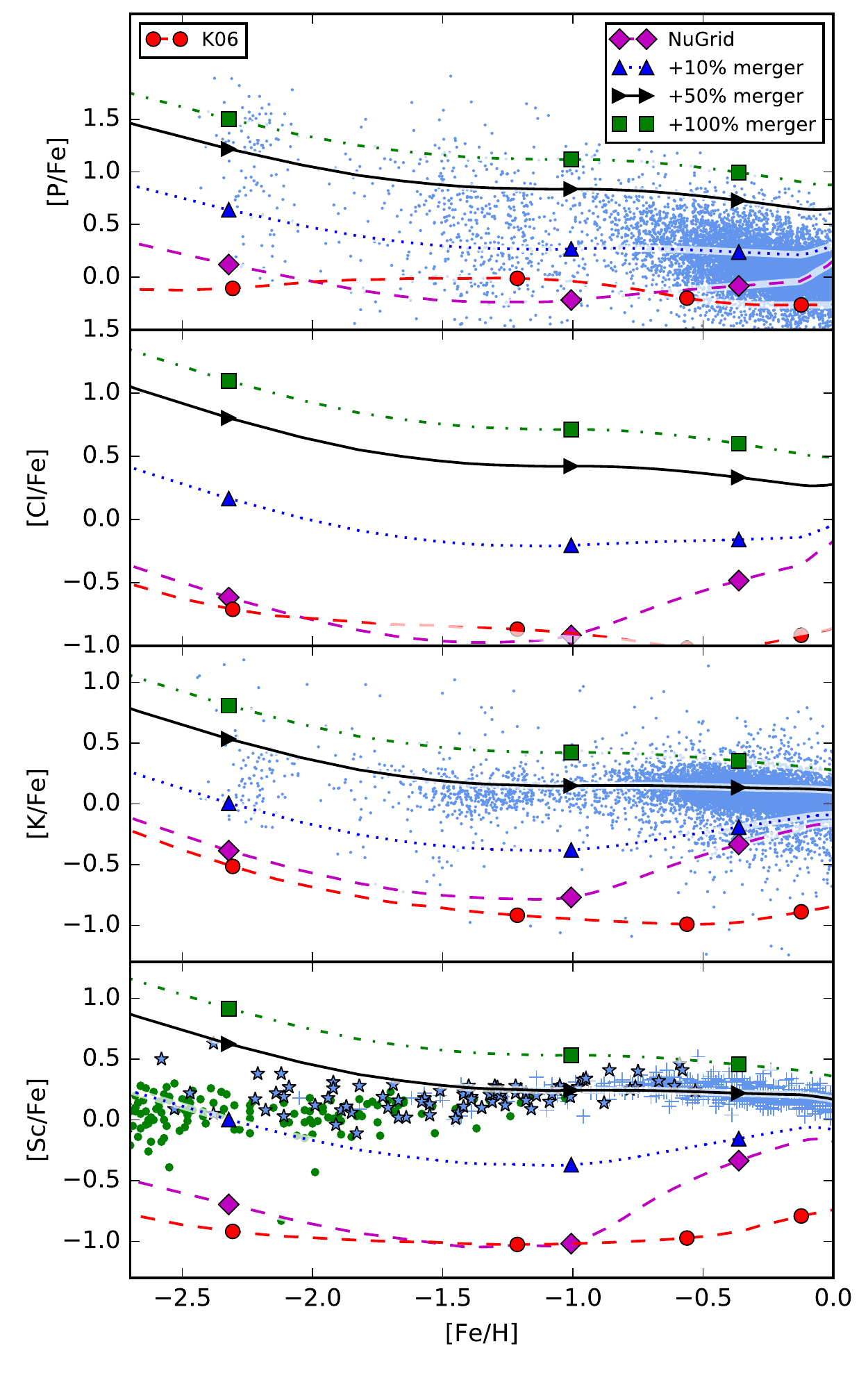}
\caption{Comparison of the predictions of P, Cl, K and Sc of our Milky
  Way model with observational data (if available).  Predictions with
  10\%, 50\% and 100\% addition of material produced in the O-C shell
  merger of the stellar model M15Z0.02 to all massive stars without
  O-C shell merger. For comparison, we show GCE predictions based on yields from K06.
  P and K data are from the APOGEE survey \citep{wilson:10,almeida:16},
 and Sc data from \citet[crosses]{ishigaki:12,ishigaki:13}, \citet[dots]{roederer:14} and \citet[stars]{battistini:15}.}
\label{fig:gceaddedmaterial}
\end{figure}

Some stellar models of R17 at $Z=0.01$ and $Z=0.02$ show O-C shell mergers (Table \ref{tab:runs}).  
These shell mergers cause the rise in [K,Sc/Fe] above
$\mathrm{[Fe/H]} > -1$ with NuGrid yields in \fig{fig:gceaddedmaterial}. We
infer that the K06 and \cite{nomoto:13} yields do not have significant O-C shell
mergers.

NuGrid models from R17 at $Z<0.01$ show convective O and C shells separated only by a thin radiative layer which
might be overcome in the real 3D star, as suggested by the convective boundary mixing reported in  3D hydrodynamic simulations (J17). Dedicated 3D hydrodynamic similations will have to determine under which conditions O-C shell mergers occur in real stars. 

We explore the assumption that shell mergers can happen in all massing star models at all Z by applying the material produced in the O-C shell merger of model M15Z0.02 to all massive star models which do not experience O-C shell mergers. We add the material to a fraction of 10\%, 50\% and 100\% of these
stars. If the merger fraction defined in this way is between 10\% and 50\%  the observed amounts of  K and Sc can be reproduced (\fig{fig:gceaddedmaterial}).

An increased $\gamma$-process production via O-C shell mergers might also boost the Galactic $\gamma$-process contribution of massive stars.
Since \cite{rayet:95} a global underproduction of p nuclides in massive stars compared to 
what is necessary to explain the solar-system distribution is found. We speculate that the shell merger production
could enable us to match the solar distribution when assuming a wide spectrum of entrainment rates, 3D hydrodynamic merger conditions
and stellar structures. 

\section{Discussion and conclusion}\label{s.discussion} 
Our results have a number of limitations. In our GCE model test we add
material produced in one particular O-C shell merger from a 1D stellar
evolution model The 1D model prediction are uncertain because 3D
effects will likely play an important role. In addition we neglect the
dependence on initial mass and metallicity.  While more massive stars
are affected by larger fallback \citep{fryer:12}, they also inhibit
larger O and C shells which boost the $\gamma$-process production.  A
larger convective C shell enables the entrainment of more material
which boosts the production of elements such as P, Cl, and
K. 
can be transported further outwards and will be less affected by
fallback.  The time span between the onset of an O-C shell merger and
core collapse is also crucial to enable sufficient element production.

Higher temperatures in the O shells of lower mass stars might lead to
a stronger production of lighter p nuclei which are transported into
the upper part of the C shell.  
For the highest masses, such as the $25\msun$ R17 models at all metallicities, the C-shell convection is either patchy or absent (especially at lower Z), which may help or hinder ingestion of C-shell material.

More generally, stellar evolution models indicate that O-C shell mergers may happen. In some cases they do, in others the shells are just stopping short of a merger. How do these results depend on various uncertain modeling assumptions in 1D, as well as on numerical convergence criteria? But even if the stellar evolution uncertainties are addressed to the best of our ability the question of whether a merger will happen in a particular, possibly marginal, case requires sufficiently realistic 3D hydrodynamic simulations. Once it is clear under which conditions O-C shell mergers happen, 3D simulations will be required to determine how O-C shell mergers proceed, whether they are ultimately dominated by non-radial, global oscillations, and how the nuclear burning energy feedback alters the flow. 

In our 3D hydrodynamic simulations, we observe approximately
spherically symmetric, quasi-steady-state behaviour for entrainment
rates that lead to noticeable odd-Z element production.
However, it appears that to fully explain the lack of Sc and K
in present yields for GCE models actual mergers of the O and C shells
are required. Our preliminary 3D simulation experiments suggest 
that the assumption of spherical symmetry may break down for such
conditions. GCE simulations which assume that O-C shell mergers as in model
M15Z0.02 occur in $\approx 10\%$ to $50\%$ of all massive stars can
account for the observed abundance of K and Sc in Milky Way
stars.

Our 1D nucleosynthesis models so far display a wide range of
\isotope[39]{K}/\isotope[41]{K} and \isotope[35]{Cl}/\isotope[37]{Cl}
ratios, depending on mixing and thermodynamic details, some of them
within a factor of a few and some much larger compared to the solar
ratios. The isotopic ratios will be an important constraint for more
realistic 3D models with updated nuclear physics.


With the entrainment of C-shell material heavy-element seeds are swept into the O shell and serve as a boost for the p-nuclei production.
In 1D stellar models we find a variety of production efficiencies which vary with initial mass and metallicity and the overproduction factors are more than 1 dex for many isotopes beyond Ba.
Such strong production might influence the GCE of p-process isotopes. A future investigation will be required to test this scenario.

\section*{Acknowledgements}
NuGrid acknowledges support from NSF grant PHY-1430152 (JINA Center
for the Evolution of the Elements).  NCSA's Blue Waters and Westgrid
provided the computing and data processing resources for this project.
RA acknowledges CITA national fellow support. BC acknowledges supported by the
FRQNT (Quebec, Canada) PDF program.  PRW
acknowledges NSF grants 1413548 and 1515792.  FH acknowledges support
from a NSERC Discovery grant.  MP acknowledges the support from SNF
(Switzerland).  SJ is a fellow of the Alexander von Humboldt
Foundation and acknowledges support from the Klaus Tschira
Stiftung. We acknowledge Brad Gibson for reminding us to check the
isotopic ratios of our models.

\bibliographystyle{apj}
\bibliography{paper}

\end{document}